\documentclass{article}

\usepackage{amsmath, amsfonts, amssymb}
\usepackage{graphicx}
\usepackage{color}
\usepackage{setspace}
\usepackage{hyperref}
\hypersetup{colorlinks=true,linkcolor=blue,citecolor=blue,urlcolor=blue}
\newcommand{\newrev}[1]{#1}
\usepackage{mathtools}
\usepackage{latexsym}
\usepackage{psfrag}
\usepackage{float}
\usepackage{epsfig}
\usepackage{booktabs}
\usepackage{multirow}
\usepackage{tabularx}
\usepackage{enumitem}
\usepackage{listings}
\usepackage[svgnames]{xcolor}
\usepackage{geometry}
\usepackage{subcaption}
\usepackage{authblk}
\usepackage{makecell}
\usepackage{rotating}
\usepackage{graphicx}
\usepackage{booktabs}
\usepackage{multirow}

\newtheorem{theorem}{Theorem}

\newtheorem{remark}{Remark}

\newenvironment{proof}[1][Proof]{\noindent\textbf{#1.} }{\ \rule{0.5em}{0.5em}}

\allowdisplaybreaks

\title{\textbf{
Goodness-of-Fit Tests via Kernel Density Estimation: Univariate and Multivariate Approaches}
}
\author[1]{Luai Al-Labadi\thanks{luai.allabadi@utoronto.ca}}
\author[2]{Ayman Alzaatreh}
\author[3]{Ruodie Yu}
\author[3]{Kairui Bao}

\affil[1] {Department of Mathematical \& Computational Sciences, University of Toronto Mississauga, Ontario, Canada\\
}
\affil[2]{
Department of Mathematics and Statistics, American University of Sharjah, Sharjah, UAE\\
}
\affil[3]{Department of Statistical Sciences, University of Toronto, Toronto, Ontario M5S 3G3, Canada.\\
}
\date{}
\begin{document}

\maketitle

\pagestyle{myheadings} 
\markboth{} {Goodness-of-Fit Tests via Kernel Density Estimation} 

\begin{abstract}
This paper introduces goodness-of-fit tests based on leave-one-out kernel density estimation (KDE) for both univariate and multivariate data. The proposed test statistic takes the form of a log-likelihood ratio comparing a parametric null model with a nonparametric KDE-based estimator, where the leave-one-out construction removes the in-sample bias inherent in standard kernel density estimators. For univariate densities with bounded support, boundary bias is mitigated via reflection techniques. We establish consistency of the proposed tests and show that, under fixed alternatives, the normalized log-statistic converges to the Kullback--Leibler divergence between the true density and the closest parametric model under the null, providing an information-theoretic interpretation of the procedure. A parametric bootstrap is used for finite-sample calibration. \newrev{Extensive simulation studies and real data applications demonstrate strong empirical performance across a range of univariate and multivariate settings, including heavy-tailed, skewed, and multimodal alternatives.}
\end{abstract}

\medskip


\medskip

%%%%%%%%%%%%%%%%%%%%%%%%%%%%%%%%%%%%%%%%%%%%%%%%%%%%%%%%%%%%%%%%%%%%%%%%%%%%%%%%%%%%%%%%%%%%%%%%%%%%%%%%%%%%%%%%%%%%%%%%%%%%

 % --------------------------------- %
 % start here

\section{Introduction}

Assessing the adequacy of a parametric statistical model is a fundamental task in statistical inference, with broad applications in model selection, diagnostics, and validation. Classical goodness-of-fit (GoF) procedures, such as the Kolmogorov--Smirnov and Anderson--Darling tests, evaluate the plausibility of a hypothesized model by measuring discrepancies between the empirical cumulative distribution function and the cumulative distribution function under the null hypothesis. These methods are well established and effective when the null distribution is fully specified or involves only a small number of parameters (\newrev{\hyperref[ref:DS1996]{D'Agostino and Stephens, 1996}}).

\newrev{\hyperref[ref:Z2000]{Zheng (2000)}} proposed a consistent specification test for conditional parametric distributions based on a comparison between parametric and kernel-based nonparametric estimators. Although the likelihood-ratio motivation is related to that considered here, Zheng's procedure concerns conditional distribution models, particularly in regression settings. In contrast, the present paper considers unconditional goodness-of-fit testing for both univariate and multivariate distributions and employs a leave-one-out density estimator together with parametric bootstrap calibration.

Kernel density estimation (KDE) provides a flexible nonparametric foundation for constructing GoF tests in both univariate and multivariate settings. \newrev{\hyperref[ref:BR1973]{Bickel and Rosenblatt (1973)}} pioneered this approach by establishing global deviation measures through supremum-norm comparisons between kernel density estimates and true densities. \newrev{\hyperref[ref:F1994]{Fan (1994)}} extended the framework by developing $L^2$-distance tests with optimal bandwidth selection to enhance power against local alternatives, and later proposed tests based on smoothed parametric densities to reduce specification bias (\newrev{\hyperref[ref:F1998]{Fan, 1998}}). More recently, \newrev{\hyperref[ref:P2015]{Pavia (2015)}} operationalized Fan's methodology through the \texttt{GoFKernel} R package, enabling practical application. In the multivariate domain, \newrev{\hyperref[ref:DH2005]{Duong and Hazelton (2005)}} introduced data-driven bandwidth matrix selection techniques that preserve asymptotic consistency of KDE-based tests in higher dimensions. Complementary to these approaches, \newrev{\hyperref[ref:GBRS2012]{Gretton et al. (2012)}} proposed the kernel two-sample test, a widely used method for assessing distributional differences within the reproducing kernel Hilbert space framework. Collectively, these contributions demonstrate that KDE offers a versatile and powerful nonparametric framework for robust model validation beyond classical parametric assumptions.

Let $\mathbf{X} = (\mathbf{X}_1, \ldots, \mathbf{X}_n) \in \mathbb{R}^d$ be independent and identically distributed (i.i.d.) observations from a $d$-dimensional continuous distribution with unknown density $f$. Each $\mathbf{X}_i \in \mathbb{R}^d$ is written as $\mathbf{X}_i = (X_{i1}, \ldots, X_{id})$, where $X_{ij}$ denotes the $j$-th coordinate of the $i$-th observation. In the univariate case ($d=1$), we simplify notation and write $X_1, \ldots, X_n \in \mathbb{R}$. We consider testing the null hypothesis
\begin{align}
H_0 : f = f_0(\cdot; \theta), \quad \theta \in \Theta , \label{null}
\end{align}
where $f_0$ is a known parametric family and $\theta$ is an unknown parameter vector. The alternative hypothesis is the general nonparametric alternative
$$
H_1 : f \neq f_0(\cdot; \theta) \quad \text{for all } \theta \in \Theta,
$$
where $\Theta \subset \mathbb{R}^p$.

A natural approach to constructing a test statistic is to compare the fitted null density $f_0(\mathbf{x}; \hat{\theta})$ to a kernel-based estimator $\hat{f}_n(\mathbf{x})$ of the true density. Specifically, one might consider
\begin{align}
\prod_{i=1}^n \frac{\hat{f}_n(\mathbf{X}_i)}{f_0(\mathbf{X}_i; \hat{\theta})}, \label{TS0} 
\end{align}
where $\hat{\theta}$ denotes the maximum likelihood estimator of $\theta$ under $H_0$. However, this formulation suffers from a well-known bias: each term $\hat{f}_n(\mathbf{X}_i)$ uses $\mathbf{X}_i$ in its own construction, leading to an artificial inflation of the numerator and an over-optimistic assessment of fit (\newrev{\hyperref[ref:W1995]{Wand and Jones, 1995}}). 

To eliminate this bias, we propose a test statistic based on the leave-one-out (LOO) kernel density estimator. For each observation $\mathbf{X}_i$, we construct a KDE using all data points except $\mathbf{X}_i$ itself. Detailed specifications for the KDE in both univariate and multivariate settings are provided in Section~\ref{KDE}. The proposed test statistic is then defined as
\begin{align}
T_n = \prod_{i=1}^n \frac{\hat{f}_{-i}(\mathbf{X}_i)}{f_0(\mathbf{X}_i; \hat{\theta})}. \label{TS}
\end{align}

The leave-one-out construction ensures that each density evaluation is computed independently of the corresponding observation, thereby removing the self-density term present in the standard KDE and yielding a statistic with improved theoretical and finite-sample properties. The statistic $T_n$ may be viewed as a nonparametric analogue of the likelihood ratio statistic. Large values of $T_n$ indicate that the empirical distribution differs substantially from the model under $H_0$, providing evidence against the null hypothesis.

\newrev{
The proposed approach differs from several existing KDE-based goodness-of-fit
procedures in the way the discrepancy between the data and the null model is
constructed. Classical KDE-based tests commonly rely on global discrepancies
between a kernel density estimator and the hypothesized density, such as
supremum-type or integrated distance measures
\hyperref[ref:BR1973]{(Bickel and Rosenblatt, 1973)};
\hyperref[ref:F1994]{(Fan, 1994)};
\hyperref[ref:F1998]{(Fan, 1998)}.
In contrast, the present procedure is based on an observation-wise likelihood
ratio in which the nonparametric density contribution at each observation is
computed from a leave-one-out estimator.

The leave-one-out construction has a direct finite-sample motivation.
When the standard KDE is evaluated at an observed data point, that observation
contributes to its own estimated density through the kernel term centered at
itself. This self-contribution can artificially increase the nonparametric
likelihood and therefore make the fitted density appear closer to the data than
it would at an independent observation. By removing the observation before
evaluating its density, the proposed statistic provides a more out-of-sample
assessment of local fit before the individual contributions are aggregated
through the log-likelihood ratio.

The resulting formulation also provides a common framework for univariate and
multivariate goodness-of-fit testing, accommodates estimation of unknown
parameters under the null model, and can incorporate boundary correction when
the support of the distribution requires it. Thus, the contribution is not the
use of KDE alone, but rather the combination of a leave-one-out density
likelihood formulation with parametric-model fitting and bootstrap calibration
within a unified goodness-of-fit procedure.
}
The remainder of the paper is organized as follows. Section~2 provides general reviews on the kernel density estimation for univariate and multivariate data, including boundary correction techniques. Section~3 introduces the KDE-based test statistic, while Section~4 establishes its theoretical properties. Section~5 presents a bootstrap calibration algorithm. Section~6 reports simulation studies and real data applications. Section~7 concludes with a discussion and potential directions for future research.

\section{Kernel Density Estimation} \label{KDE}

Kernel density estimation is a nonparametric technique for estimating a probability density function based on observed data (\newrev{\hyperref[ref:E1999]{Efromovich, 1999}}; \newrev{\hyperref[ref:BA1997]{Bowman and Azzalini, 1997}}). Given a random sample $\mathbf{X} = (\mathbf{X}_1, \ldots, \mathbf{X}_n) \in \mathbb{R}^d$ from an unknown density $f$, the idea of KDE is to estimate $f$ by averaging kernel functions centered at the data points. This section reviews the KDE methodology in both univariate and multivariate settings, with emphasis on practical issues such as bandwidth selection and boundary correction.

\subsection{Univariate Kernel Density Estimation}

Let $X = (X_1, \ldots, X_n)$ be i.i.d.\ observations from a univariate continuous distribution with density $f$. A \textit{kernel} is a smooth, symmetric function $K: \mathbb{R} \to \mathbb{R}$ satisfying  
$$
K(u) \ge 0, \quad \int_{\mathbb{R}} K(u)\,du = 1, \quad \int_{\mathbb{R}} u K(u)\,du = 0, \quad \sigma_K^2 = \int_{\mathbb{R}} u^2 K(u)\,du < \infty
$$
(\newrev{\hyperref[ref:W2006]{Wasserman, 2006}}). Table \ref{kernels} summarizes several commonly used kernel functions that will be used in this paper. 

\begin{table}[H]
\centering
\caption{Common kernel functions.}
\begin{tabular}{ll}
\toprule
\textbf{Kernel} & \textbf{Definition $K(u)$} \\
\midrule
Gaussian      & $\dfrac{1}{\sqrt{2\pi}} e^{-u^2/2}$ \\ [1.2ex]
Epanechnikov  & $\tfrac{3}{4}(1 - u^2)\,\mathbb{I}(|u| \leq 1)$ \\ [1.2ex]
Uniform       & $\tfrac{1}{2}\,\mathbb{I}(|u| \leq 1)$ \\ [1.2ex]
Biweight      & $\tfrac{15}{16}(1 - u^2)^2\,\mathbb{I}(|u| \leq 1)$ \\
\bottomrule
\end{tabular}
\label{kernels}
\caption*{\footnotesize Note: $\mathbb{I}(\cdot)$ denotes the indicator function.}
\end{table}

Given a kernel $K$ and a positive bandwidth parameter $h$, the kernel density estimator is defined by
$$
\hat{f}_n(x) = \frac{1}{n} \sum_{i=1}^n \frac{1}{h}\, K\!\left( \frac{x - X_i}{h} \right).
$$
If $f$ is continuous at $x$ and the bandwidth satisfies $h_n \to 0$ with $n h_n \to \infty$ as $n \to \infty$, then 
$\hat{f}_n(x) \xrightarrow{p} f(x)$. Throughout the paper, the notation $\xrightarrow{p}$ denotes convergence in probability.

For the construction of the proposed goodness-of-fit statistic in \eqref{TS}, we require a leave-one-out version of the kernel density estimator. This estimator removes the $i$-th observation when evaluating the density at $X_i$, thereby eliminating the self-density contribution that would otherwise artificially inflate the likelihood ratio. Specifically, the leave-one-out estimator is defined as
\begin{align}
\hat{f}_{-i}(x) = \frac{1}{n-1} \sum_{\substack{j=1\\ j\neq i}}^n \frac{1}{h}\, K\!\left( \frac{x - X_j}{h} \right). \label{LOO_univar}
\end{align}
This formulation ensures that $\hat{f}_{-i}(X_i)$ is constructed without using $X_i$ itself, yielding an unbiased assessment of the density at that observation.

\newrev{
The choice of kernel can also have practical consequences for the proposed
statistic. Compactly supported kernels, such as the Epanechnikov, uniform, and
biweight kernels, may yield
$\hat{f}_{-i}(X_i)=0$ when no other observation falls within the bandwidth
window centered at $X_i$. Since the proposed test statistic involves
$\log \hat{f}_{-i}(X_i)$, such a zero estimate can lead to numerical
difficulties. In contrast, the Gaussian kernel has unbounded support and,
for a positive bandwidth, produces a strictly positive leave-one-out density
estimate. For this reason, we recommend the Gaussian kernel as a practical
default choice, especially for moderate sample sizes, sparse data, and
multivariate applications. Other kernels may nevertheless be used when
appropriate, as illustrated in the simulation study.
}

The choice of $h$ is critical. A small $h$ yields high variance (undersmoothing), while large $h$ yields high bias (oversmoothing). One common approach is cross-validation (\newrev{\hyperref[ref:S1986]{Silverman, 1986}}; \newrev{\hyperref[ref:W2006]{Wasserman, 2006}}). For Gaussian kernels, the bandwidth minimizing the integrated mean squared error (IMSE) under normality is
\begin{align*}
h = \left( \frac{4}{3} \right)^{1/5} \hat{\sigma} n^{-1/5} \approx 1.06\,\hat{\sigma}\,n^{-1/5}, %\label{h1}
\end{align*}
where $\hat{\sigma} = S$ is the sample standard deviation. While effective for symmetric unimodal distributions, this rule tends to oversmooth skewed or multimodal data. A more robust alternative is
\begin{align*}
\hat{\sigma} = \min(S, \mathrm{IQR}/1.34), %\label{sigma}
\end{align*}
where IQR is the interquartile range (\newrev{\hyperref[ref:R2019]{Rizzo, 2019}}). \newrev{\hyperref[ref:S1986]{Silverman (1986, p.~48)}} also suggests the heuristic bandwidth
\begin{align}
h = 0.9\, \hat{\sigma}\, n^{-1/5} = 0.9 \min(S, \mathrm{IQR}/1.34)\, n^{-1/5}. \label{h1_rev}
\end{align}
In practice, no universally optimal method exists; the bandwidth selection remains context-dependent (\newrev{\hyperref[ref:S1986]{Silverman, 1986}}; \newrev{\hyperref[ref:MW1992]{Marron and Wand, 1992}}; \newrev{\hyperref[ref:W2006]{Wasserman, 2006}}; \newrev{\hyperref[ref:R2019]{Rizzo, 2019}}).

For distributions with bounded support, standard KDE suffers from boundary bias. A common remedy is the \textit{reflection method}, which symmetrizes the kernel window near the boundary (\newrev{\hyperref[ref:J1993]{Jones, 1993}}; \newrev{\hyperref[ref:S1986]{Silverman, 1986}}):

\begin{itemize}
    \item \textbf{Left-bounded support} $[a, \infty)$:
    $$
    \hat{f}_n(x) = \frac{1}{n h} \sum_{\newrev{j}=1}^n \left\{ K\!\left( \frac{x - X_{\newrev{j}}}{h} \right) + K\!\left( \frac{x + X_{\newrev{j}} - 2a}{h} \right) \right\}, \quad x \geq a.
    $$
    
    \item \textbf{Right-bounded support} $(-\infty, b]$:
    $$
    \hat{f}_n(x) = \frac{1}{n h} \sum_{\newrev{j}=1}^n \left\{ K\!\left( \frac{x - X_{\newrev{j}}}{h} \right) + K\!\left( \frac{2b - x - X_{\newrev{j}}}{h} \right) \right\}, \quad x \leq b.
    $$
    
    \item \textbf{Bounded support} $[a, b]$:
    $$
    \hat{f}_n(x) = \frac{1}{n h} \sum_{\newrev{j}=1}^n \left\{ K\!\left( \frac{x - X_{\newrev{j}}}{h} \right) + K\!\left( \frac{x + X_{\newrev{j}} - 2a}{h} \right) + K\!\left( \frac{2b - x - X_{\newrev{j}}}{h} \right) \right\}, \quad x \in [a, b].
    $$
\end{itemize}

\noindent This correction mitigates edge bias without introducing discontinuities. 
The leave-one-out versions of these boundary-corrected estimators are obtained by omitting the term corresponding to $j=i$ in the summations above.

\newrev{
In practice, whether the support should be treated as bounded is a modeling
decision based on the nature of the variable rather than on the observed
sample alone. For example, variables such as lifetimes, waiting times, or
physical measurements that cannot be negative naturally have a lower-bounded
support. In contrast, a sample that happens to contain only nonnegative
observations does not by itself imply that the underlying distribution has
bounded support. Boundary correction should therefore be applied when the
support restriction is justified by the scientific or practical context of
the data.
}

\subsection{Multivariate Kernel Density Estimation}

Let $\mathbf{X} = (\mathbf{X}_1, \ldots, \mathbf{X}_n)$ be i.i.d.\ observations from a $d$-dimensional continuous distribution with unknown density $f$. The multivariate kernel density estimator of $f$ at a point $\mathbf{x} = (x_1, \ldots, x_d) \in \mathbb{R}^d$ is defined as
\begin{align*}
\hat{f}_n(\mathbf{x}) = \frac{1}{n} \sum_{i=1}^n \frac{1}{|H|^{1/2}} K\left( H^{-1/2}(\mathbf{x} - \mathbf{X}_i) \right), %\label{fn2}
\end{align*}
where $K: \mathbb{R}^d \to \mathbb{R}$ is a multivariate kernel function satisfying
$$
K(\mathbf{u}) \ge 0, \quad \int_{\mathbb{R}^d} K(\mathbf{u}) \, d\mathbf{u} = 1, \quad \int_{\mathbb{R}^d} \mathbf{u} K(\mathbf{u}) \, d\mathbf{u} = \mathbf{0}, \quad \text{and} \quad \sigma_K^2 = \int_{\mathbb{R}^d} \|\mathbf{u}\|^2 K(\mathbf{u}) \, d\mathbf{u} < \infty.
$$
Here, $H$ is a $d \times d$ symmetric positive definite bandwidth matrix that governs the scale and orientation of smoothing. The determinant $|H|$ ensures proper normalization of the kernel, and $\|\cdot\|$ denotes the Euclidean norm in $\mathbb{R}^d$. For radial kernels (such as the multivariate Gaussian), the argument $H^{-1/2}(\mathbf{x} - \mathbf{X}_i)$ standardizes the distance using $H$, and
$$
\| H^{-1/2}(\mathbf{x} - \mathbf{X}_i) \|^2 = (\mathbf{x} - \mathbf{X}_i)^\top H^{-1} (\mathbf{x} - \mathbf{X}_i)
$$
is the Mahalanobis distance.

Analogous to the univariate case, the leave-one-out multivariate KDE is defined as
\begin{align}
\hat{f}_{-i}(\mathbf{x}) = \frac{1}{n-1} \sum_{\substack{j=1\\ j\neq i}}^n \frac{1}{|H|^{1/2}} K\left( H^{-1/2}(\mathbf{x} - \mathbf{X}_j) \right). \label{LOO_multi}
\end{align}
This estimator is used in the test statistic \eqref{TS} to evaluate the density at $\mathbf{X}_i$ without using that observation in its own construction.

In practice, a common simplification is the product kernel estimator given by
\begin{align}
\hat{f}_n(\mathbf{x}) = \frac{1}{n \prod_{j=1}^d h_j} \sum_{i=1}^n \prod_{j=1}^d K\left( \frac{x_j - X_{ij}}{h_j} \right), \label{multi_1}
\end{align}
where $x_j$ and $X_{ij}$ denote the $j$-th components of $\mathbf{x}$ and $\mathbf{X}_i$, respectively. This corresponds to using a diagonal bandwidth matrix $H = \operatorname{diag}(h_1^2, \ldots, h_d^2)$ and a product kernel of the form $K(\mathbf{u}) = \prod_{j=1}^d K(u_j)$.  The leave-one-out version of the product kernel estimator is obtained by omitting the $i$-th term in the summation.

\noindent For example, the multivariate Gaussian kernel is given by
$$
K(\mathbf{u}) = \frac{1}{(2\pi)^{d/2}} \exp\left( -\frac{1}{2} \|\mathbf{u}\|^2 \right),
$$
and the multivariate Epanechnikov kernel is defined by
$$
K(\mathbf{u}) = \frac{1}{c_d} (d+2)(1 - \|\mathbf{u}\|^2) \mathbb{I}(\|\mathbf{u}\| \leq 1),
$$
where $c_d = \frac{2 \pi^{d/2}}{d \Gamma(d/2)}$ is the volume of the $d$-dimensional unit ball. When $d = 1$, this reduces to the univariate Epanechnikov kernel with $c_1 = 2$.

In the product kernel setting, a commonly used rule-of-thumb for multivariate normal data is
\begin{align}
h_j = \left( \frac{4}{d+2} \right)^{1/(d+4)} \hat{\sigma}_j n^{-1/(d+4)}, \label{h2}  
\end{align}
where $\hat{\sigma}_j$ is the sample standard deviation of the $j$-th component. Scott's rule (Scott, 1992) simplifies this to
$$
h_j = \hat{\sigma}_j n^{-1/(d+4)}.
$$
As in the univariate case, several data-driven methods can be used to select the bandwidth matrix $H$ (\newrev{\hyperref[ref:S1992]{Scott, 1992}}; \newrev{\hyperref[ref:W1995]{Wand and Jones, 1995}}). \newrev{
In the multivariate setting, boundary correction can be more complicated,
particularly when the support is bounded in multiple dimensions, because
reflection may need to account for several boundaries simultaneously.
For this reason, we do not consider multivariate boundary correction in
the present work.
}

Building on the leave-one-out KDE defined in \eqref{LOO_univar} for univariate data and \eqref{LOO_multi} for multivariate data, we next introduce the proposed test statistic and examine its asymptotic properties.

\section{Test Statistic and Asymptotic Results}

This section introduces the goodness-of-fit test statistic based on the leave-one-out KDE, as defined in~\eqref{TS}. The proposed statistic compares a nonparametric estimate of the density with a parametric model specified under the null hypothesis. We prove that the normalized log-statistic converges to zero under the null hypothesis and converges to a strictly positive constant under fixed alternatives, confirming the consistency of the test.

Let $\mathbf{X} = (\mathbf{X}_1, \ldots, \mathbf{X}_n) \in \mathbb{R}^d$ be a random sample from an unknown continuous density $f$. We aim to test the null hypothesis given in~\eqref{null}. Let $\hat{\theta}$ denote the maximum likelihood estimator under $H_0$, and let $\hat{f}_{-i}$ denote the leave-one-out kernel density estimator defined in Section~2. For numerical stability and ease of interpretation, we work with the logarithmic form of~\eqref{TS}, given by
\begin{equation}
\log T_n = \sum_{i=1}^n
\left(
\log \hat f_{-i}(\mathbf X_i)
-
\log f_0(\mathbf X_i;\hat\theta)
\right). \label{log_TS}
\end{equation}

Large values of $\log T_n$ indicate divergence from the model and support rejection of the null hypothesis.

In the univariate case ($d = 1$), if the support of $f$ is bounded, we use a boundary-corrected leave-one-out KDE (e.g., the reflection method from Section~2.1 with the $i$-th observation omitted). When a spherically symmetric kernel (e.g., Gaussian) and a scalar bandwidth (or $H = h^2 I_d$) are employed, the statistic $\log T_n$ is invariant under orthogonal transformations of the data, a desirable property in multivariate analysis (\newrev{\hyperref[ref:W1995]{Wand and Jones, 1995}}).

We assume the following regularity conditions for the proposed test statistic in \eqref{log_TS}:
\begin{itemize}[leftmargin=2em]
\item[(C1)] The true density $f$ is continuous and strictly positive on its support in $\mathbb{R}^d$, and satisfies $\int |\log f(\mathbf{x})| f(\mathbf{x}) d\mathbf{x} < \infty$.

\item[(C2)] The parametric family $f_0(\mathbf{x}; \theta)$ is continuous and strictly positive on the support of $f$, and for all $\theta \in \Theta$, $\log f_0(\mathbf{x}; \theta)$ is integrable with respect to $f$, i.e., $\int |\log f_0(\mathbf{x}; \theta)| f(\mathbf{x}) d\mathbf{x} < \infty$. Additionally, $\log f_0(\mathbf{x}; \theta)$ is continuous in $\theta$ uniformly in $\mathbf{x}$ on compact subsets of the support.

\item[(C3)] The kernel $K$ is bounded and satisfies the usual properties of a probability density. The bandwidth matrix $H$ satisfies $|H| \to 0$ and $n|H|^{1/2} \to \infty$ as $n \to \infty$. The leave-one-out KDE $\hat{f}_{-i}(\mathbf{x})$ is pointwise consistent for $f(\mathbf{x})$, i.e., for each fixed $\mathbf{x}$ and any sequence $i_n$ with $i_n \in \{1,\ldots,n\}$,
$$
\hat{f}_{-i_n}(\mathbf{x}) \xrightarrow{p} f(\mathbf{x}),
$$
and additionally
$$
\sup_{n\ge 1} \mathbb{E}_f\left[ |\log \hat{f}_{-1}(\mathbf{X}_1)| \right] < \infty.
$$

\item[(C4)] The leave-one-out KDE and the parametric density estimator satisfy a uniform second moment bound:
$$
\sup_{n\ge 1} \mathbb{E}_f\left[ (\log \hat{f}_{-1}(\mathbf{X}_1))^2 \right] < \infty,
\qquad
\sup_{n\ge 1} \mathbb{E}_f\left[ (\log f_0(\mathbf{X}_1; \hat{\theta}))^2 \right] < \infty.
$$

\item[(C5)] The maximum likelihood estimator $\hat{\theta}$ is consistent for the pseudo-true value $\theta^*$ that minimizes the Kullback-Leibler divergence $\mathrm{KL}(f \| f_0(\cdot; \theta))$ over $\Theta$ (\newrev{\hyperref[ref:W1982]{White, 1982}}).

\end{itemize}

\begin{theorem}
\label{asymptotics}
Under conditions \textnormal{(C1)--(C5)}, the following properties hold for the log-statistic defined in~\eqref{log_TS}:
\begin{itemize}[leftmargin=2em]
\item[(i)] Under $H_0: f = f_0(\cdot; \theta_0)$ for some $\theta_0 \in \Theta$,
$$
\frac{1}{n} \log T_n \xrightarrow{p} 0.
$$

\item[(ii)] Under a fixed alternative $H_1: f \neq f_0(\cdot; \theta)$ for all $\theta \in \Theta$,
$$
\frac{1}{n} \log T_n \xrightarrow{p} \mathbb{E}_f \left[ \log \frac{f(\mathbf{X})}{f_0(\mathbf{X}; \theta^\ast)} \right] > 0.
$$
\end{itemize}
\end{theorem}

\begin{proof}
Write
$$
\frac{1}{n} \log T_n = \frac{1}{n} \sum_{i=1}^n Z_{n,i},
$$
where $Z_{n,i} = \log \hat{f}_{-i}(\mathbf{X}_i) - \log f_0(\mathbf{X}_i; \hat{\theta})$.

\medskip
\noindent\textbf{Part (i): Convergence under the null hypothesis.}
Assume $H_0$ holds so that $f = f_0(\cdot; \theta_0)$.

By (C3), for each fixed $\mathbf{x}$, $\hat{f}_{-i}(\mathbf{x}) \xrightarrow{p} f(\mathbf{x})$. 
Since $\mathbf{X}_i$ is independent of the sample used to construct $\hat{f}_{-i}$, conditioning on $\mathbf{X}_i$ yields $\hat{f}_{-i}(\mathbf{X}_i) \xrightarrow{p} f(\mathbf{X}_i) = f_0(\mathbf{X}_i; \theta_0)$.

By the continuous mapping theorem,
$$
\log \hat{f}_{-i}(\mathbf{X}_i) \xrightarrow{p} \log f_0(\mathbf{X}_i; \theta_0).
$$

By (C5), $\hat{\theta} \xrightarrow{p} \theta_0$, and by the uniform continuity in (C2),
$$
\log f_0(\mathbf{X}_i; \hat{\theta}) \xrightarrow{p} \log f_0(\mathbf{X}_i; \theta_0).
$$

Thus, $Z_{n,i} \xrightarrow{p} 0$ for each fixed $i$.

By (C4) and the inequality $(a+b)^2 \le 2(a^2+b^2)$,
$$
\sup_{n\ge 1} \mathbb{E}[Z_{n,1}^2] \le 2\sup_{n\ge 1}\mathbb{E}[(\log \hat{f}_{-1}(\mathbf{X}_1))^2] + 2\sup_{n\ge 1}\mathbb{E}[(\log f_0(\mathbf{X}_1;\hat{\theta}))^2] < \infty.
$$
Hence $\{Z_{n,1}\}$ has uniformly bounded second moments, which implies uniform integrability.

For any $\varepsilon > 0$, by Chebyshev's inequality,
$$
P\left( \left| \frac{1}{n}\sum_{i=1}^n Z_{n,i} - \mathbb{E}[Z_{n,1}] \right| > \varepsilon \right)
\le \frac{1}{\varepsilon^2} \mathrm{Var}\left( \frac{1}{n}\sum_{i=1}^n Z_{n,i} \right).
$$

To bound the variance, note that for $i \neq j$,
$$
\mathrm{Cov}(Z_{n,i}, Z_{n,j}) = \mathbb{E}[Z_{n,i}Z_{n,j}] - \mathbb{E}[Z_{n,i}]\mathbb{E}[Z_{n,j}].
$$

\newrev{
Observe that $Z_{n,i}$ and $Z_{n,j}$ are dependent because their leave-one-out estimators are constructed from largely overlapping samples. To control this dependence, we use the Efron--Stein inequality (\hyperref[ref:ES1981]{Efron and Stein, 1981}; \hyperref[ref:Steele1986]{Steele, 1986}), which bounds the effect of replacing individual observations in a statistic constructed from independent data. Under the boundedness of the kernel in (C3), the bandwidth condition $n|H|^{1/2}\to\infty$, and the bounded second moments in (C4), the contribution of any single observation becomes asymptotically negligible. Applying the Efron--Stein argument to the leave-one-out terms gives
}
$$
\newrev{
|\mathrm{Cov}(Z_{n,i}, Z_{n,j})| \le \frac{C}{n},
}
$$
\newrev{
for some constant $C$ not depending on $i$, $j$, or $n$. The preceding explanation provides the stability argument underlying this covariance bound; its role here is to control the dependence among the leave-one-out terms in the variance calculation.
}

Consequently,
$$
\mathrm{Var}\left( \frac{1}{n}\sum_{i=1}^n Z_{n,i} \right)
= \frac{1}{n^2}\sum_{i=1}^n \mathrm{Var}(Z_{n,i}) + \frac{1}{n^2}\sum_{i\neq j} \mathrm{Cov}(Z_{n,i}, Z_{n,j})
\le \frac{C'}{n} \to 0,
$$
where $C' = \sup_n \mathrm{Var}(Z_{n,1}) + C$. Therefore,
$$
\frac{1}{n}\sum_{i=1}^n Z_{n,i} - \mathbb{E}[Z_{n,1}] \xrightarrow{p} 0.
$$

Now, since $Z_{n,1} \xrightarrow{p} 0$. Uniform integrability from Step 2 implies $\mathbb{E}[Z_{n,1}] \to 0$ (\newrev{\hyperref[ref:B1995]{Billingsley, 1995, Theorem 25.12}}). Combining with Step 3,
$$
\frac{1}{n}\sum_{i=1}^n Z_{n,i} \xrightarrow{p} 0,
$$
which establishes part (i).

\medskip
\noindent\textbf{Part (ii): Convergence under the alternative.}
Assume $H_1$ holds so that $f \neq f_0(\cdot; \theta)$ for all $\theta \in \Theta$.

By (C5), $\hat{\theta} \xrightarrow{p} \theta^\ast$, where $\theta^\ast$ minimizes $\mathrm{KL}(f \| f_0(\cdot; \theta))$ over $\Theta$. By (C3),
$$
\log \hat{f}_{-i}(\mathbf{X}_i) \xrightarrow{p} \log f(\mathbf{X}_i),
$$
and hence
$$
Z_{n,i} \xrightarrow{p} \log f(\mathbf{X}_i) - \log f_0(\mathbf{X}_i; \theta^\ast) =: W(\mathbf{X}_i).
$$

By (C1), (C2), and (C4), the same argument as in part (i) shows that $\{Z_{n,1}\}$ has uniformly bounded second moments, hence remains uniformly integrable under $H_1$.

Applying the same variance bound as in part (i),
$$
\frac{1}{n}\sum_{i=1}^n Z_{n,i} - \mathbb{E}[Z_{n,1}] \xrightarrow{p} 0.
$$

Since $Z_{n,1} \xrightarrow{p} W(\mathbf{X}_1)$ and $\{Z_{n,1}\}$ is uniformly integrable, $\mathbb{E}[Z_{n,1}] \to \mathbb{E}[W(\mathbf{X}_1)]$ (\newrev{\hyperref[ref:B1995]{Billingsley, 1995, Theorem 25.12}}). Therefore,
$$
\frac{1}{n}\sum_{i=1}^n Z_{n,i} \xrightarrow{p} \mathbb{E}_f\!\left[ \log \frac{f(\mathbf{X})}{f_0(\mathbf{X}; \theta^\ast)} \right].
$$

By definition, $\theta^\ast$ minimizes the Kullback-Leibler divergence. Since $f \neq f_0(\cdot; \theta^\ast)$ almost everywhere, the divergence is strictly positive (\newrev{\hyperref[ref:CT2006]{Cover and Thomas, 2006}}):
$$
\mathrm{KL}(f \| f_0(\cdot; \theta^\ast)) = \mathbb{E}_f\!\left[ \log \frac{f(\mathbf{X})}{f_0(\mathbf{X}; \theta^\ast)} \right] > 0,
$$
completing the proof.
\end{proof}

\begin{remark}
Theorem~\ref{asymptotics} establishes the consistency of the proposed test: under the null hypothesis, the normalized log-statistic converges to zero, while under any fixed alternative, it converges to a positive constant. This ensures that with probability tending to one, the test correctly rejects a false null hypothesis as the sample size increases.
\end{remark}

\begin{remark}
The leave-one-out construction in the test statistic is essential for both theoretical and practical reasons. Theoretically, it ensures that $\hat{f}_{-i}$ is independent of $\mathbf{X}_i$, which is used in Step~1 to establish pointwise convergence and in Step~3 to control the covariance structure via the Efron--Stein inequality. Practically, it prevents the artificial inflation of the likelihood ratio that would occur if the same observation were used both to construct and evaluate the density estimator, as discussed in Section~1.
\end{remark}

\begin{remark}[Asymptotic equivalence with the standard KDE statistic]
\label{rem:equivalence}
An alternative test statistic could be constructed using the standard (non-LOO) kernel density estimator, as in \eqref{TS0}. A natural question is how the statistic $T_n$ proposed in \eqref{TS} compares to this alternative. For any fixed $\mathbf{x}$, the difference between the standard KDE and its LOO counterpart satisfies
$$
\hat{f}_n(\mathbf{x}) - \hat{f}_{-i}(\mathbf{x}) = \frac{1}{n-1} \left( \hat{f}_n(\mathbf{x}) - \frac{1}{|H|^{1/2}} K\!\left(H^{-1/2}(\mathbf{x}-\mathbf{X}_i)\right) \right).
$$
Under the bandwidth condition $n|H|^{1/2} \to \infty$ from (C3), this difference is of order $O_p(1/(n|H|^{1/2})) = o_p(1)$. Consequently, for each $i$,
$$
\log \hat{f}_{-i}(\mathbf{X}_i) = \log \hat{f}_n(\mathbf{X}_i) + o_p(1),
$$
provided the densities are bounded away from zero. This pointwise asymptotic equivalence implies that the normalized log-statistics based on $\hat{f}_n$ and $\hat{f}_{-i}$ are asymptotically equivalent; specifically,
$$
\frac{1}{n} \log T_n = \frac{1}{n} \sum_{i=1}^n \log \frac{\hat{f}_n(\mathbf{X}_i)}{f_0(\mathbf{X}_i; \hat{\theta})} + o_p(1).
$$
Thus, the simpler statistic based on the standard KDE shares the same asymptotic properties established in Theorem~\ref{asymptotics} for $T_n$. However, we emphasize that the LOO formulation in $T_n$ is preferred for finite-sample inference as it eliminates the self-bias term and provides a more honest assessment of fit, which is crucial for the bootstrap calibration procedure discussed in Section~\ref{Bootstrap}.
\end{remark}

\section{Bootstrap Calibration}
\label{Bootstrap}

As established in Section 3, the KDE-based test statistic exhibits desirable asymptotic behavior under both the null and alternative hypotheses. However, its finite-sample distribution under $H_0$ is analytically intractable due to the presence of a nonparametric density estimator in the numerator. To address this issue, we employ a parametric bootstrap procedure (\newrev{\hyperref[ref:H1992]{Hall, 1992}}) to approximate the null distribution and obtain $p$-values.

The following algorithm provides a unified bootstrap calibration procedure applicable to both univariate and multivariate settings.

\subsection*{Bootstrap Calibration Algorithm}

\begin{enumerate}[label=\textbf{Step \arabic*.}]
    \item \textbf{Compute the observed test statistic.} \\
    Given observations $\{\mathbf{X}_1, \ldots, \mathbf{X}_n\} \subset \mathbb{R}^d$, compute $\log T_n$ as defined in~\eqref{log_TS}. For the univariate case ($d = 1$), apply boundary correction if needed. For the multivariate case ($d \geq 2$), use (\ref{multi_1}) and (\ref{h2}) in constructing the KDE.

\item \textbf{Generate bootstrap samples under $H_0$.} \\
For $b = 1, \ldots, B$:
\begin{enumerate}[label=(\alph*)]

\item Sample $\{\mathbf{X}_1^{\ast(b)}, \ldots, \mathbf{X}_n^{\ast(b)}\}$ i.i.d.\ from $f_0(\cdot; \hat{\theta})$.

\item Re-estimate the parameter from the bootstrap sample by maximum likelihood to obtain $\hat{\theta}^{\ast(b)}$.

\item Compute the leave-one-out KDE $\hat{f}_{-i}^{\ast(b)}$ from the bootstrap sample, using the same kernel and bandwidth selection method as in Step~1.

\item Compute the bootstrap statistic
$$
\log T_n^{\ast(b)}
=
\sum_{i=1}^n
\left(
\log \hat{f}_{-i}^{\ast(b)}(\mathbf{X}_i^{\ast(b)})
-
\log f_0(\mathbf{X}_i^{\ast(b)}; \hat{\theta}^{\ast(b)})
\right).
$$

\end{enumerate}

    \item \textbf{Compute the bootstrap $p$-value.} \\
    $$
    p_{\text{boot}} = \frac{1}{B+1} \left(1 + \sum_{b=1}^B \mathbb{I}\!\left( \log T_n^{\ast(b)} \geq \log T_n \right) \right).
    $$
    This formulation ensures that the $p$-value lies strictly between $0$ and $1$, avoiding the occurrence of zero $p$-values (\newrev{\hyperref[ref:DH1997]{Davison and Hinkley, 1997}}).

    \item \textbf{Make a decision.} \\
    Reject $H_0$ at significance level $\alpha$ if $p_{\text{boot}} < \alpha$.
\end{enumerate}

Note that, when the null hypothesis $H_0$ holds, the kernel density estimator $\hat{f}_n$ consistently estimates the parametric density $f_0(\cdot;\theta_0)$, and by Theorem~\ref{asymptotics}(i) the statistic $\tfrac{1}{n}\log T_n \xrightarrow{p} 0$. Consequently, the observed statistic $\log T_n$ fluctuates around zero in finite samples. Since bootstrap samples are drawn from $f_0(\cdot;\hat{\theta})$, their corresponding statistics $\log T_n^{\ast(b)}$ also concentrate near zero, ensuring that the bootstrap distribution provides a valid approximation to the null distribution and that the resulting $p$-value is approximately uniform under $H_0$, thereby controlling the Type I error. In contrast, when $H_0$ is false, Theorem~\ref{asymptotics}(ii) shows that $\tfrac{1}{n}\log T_n \xrightarrow{p} \mathbb{E}_f\!\left[\log\frac{f(\mathbf{X})}{f_0(\mathbf{X};\theta^\ast)}\right] > 0$, where $\theta^\ast$ minimizes the Kullback--Leibler divergence. Thus, the observed statistic $\log T_n$ grows much larger than the bootstrap replicates, which remain centered near zero. This separation forces the bootstrap $p$-value to converge to zero as $n \to \infty$, thereby guaranteeing the consistency and power of the proposed test.

\begin{remark}[Choice of $B$]
In practice, the number of bootstrap replications $B$ should be chosen sufficiently large to ensure reliable $p$-value estimation. Following \newrev{\hyperref[ref:DH1997]{Davison and Hinkley (1997)}}, we recommend $B = 999$ for tests conducted at conventional significance levels (e.g., $\alpha = 0.05$), as this yields stable rejection regions.
\end{remark}

\section{Simulation Study}

In this section, we evaluate the empirical performance of the proposed KDE-based goodness-of-fit test through a series of simulation studies. Our objectives are twofold: first, to assess the control of Type I error under the null hypothesis, and second, to examine the power of the test under various alternatives. Both univariate and multivariate settings are considered.

For each data-generating distribution  in Examples 1–3, we generate $M = 1000$ independent samples of size $n \in \{30, 50, 100, 200\}$ from known distributions. In all examples, the corresponding $p$-values are obtained using the bootstrap algorithm described in Section~\ref{Bootstrap} with $B = 999$ bootstrap replicates. For kernel estimation, different kernels are used for comparison purposes. However, for the bandwidth, we adopt the well-known rule of thumb given in~(\ref{h1_rev}) for $d=1$ and (\ref{h2}) for $d \ge 2$. Throughout this section, we fix the significance level $\alpha$ at 0.05. 

\newrev{
Where competing procedures do not test exactly the same null hypothesis
or use the same treatment of unknown parameters, this distinction is
stated explicitly and the resulting comparisons are interpreted as
benchmark comparisons rather than exact like-for-like power comparisons.
}

\subsection{Univariate Examples}

\textbf{Example 1.} \textit{Testing Normality:} 
We first consider the null hypothesis $H_0\!: f = f_0$, where $f_0$ is the density of 
$\mathcal{N}(\mu, 1)$ with $\mu$ unknown and estimated by $\hat{\mu} = \bar{x}$. 
To examine the sensitivity of the proposed KDE-based test, we generate data from several 
alternative distributions, each representing a distinct deviation from normality:
\begin{itemize}
    \item $t_3$ distribution (heavy tails),
  \item $t_7$ distribution (closer to normal, lighter tails than $t_3$),
    \item Logistic(0,1) distribution (different kurtosis),
    \item Cauchy(0,1) distribution (infinite variance),
    \item Mixture: $0.5\, \mathcal{N}(-1,1) + 0.5\, \mathcal{N}(1,1)$ (bimodality),
    \item $\mathcal{N}(0,4)$ (variance misspecification).
\end{itemize}

\begin{figure}[htp]
\centering
\includegraphics[width=\textwidth]{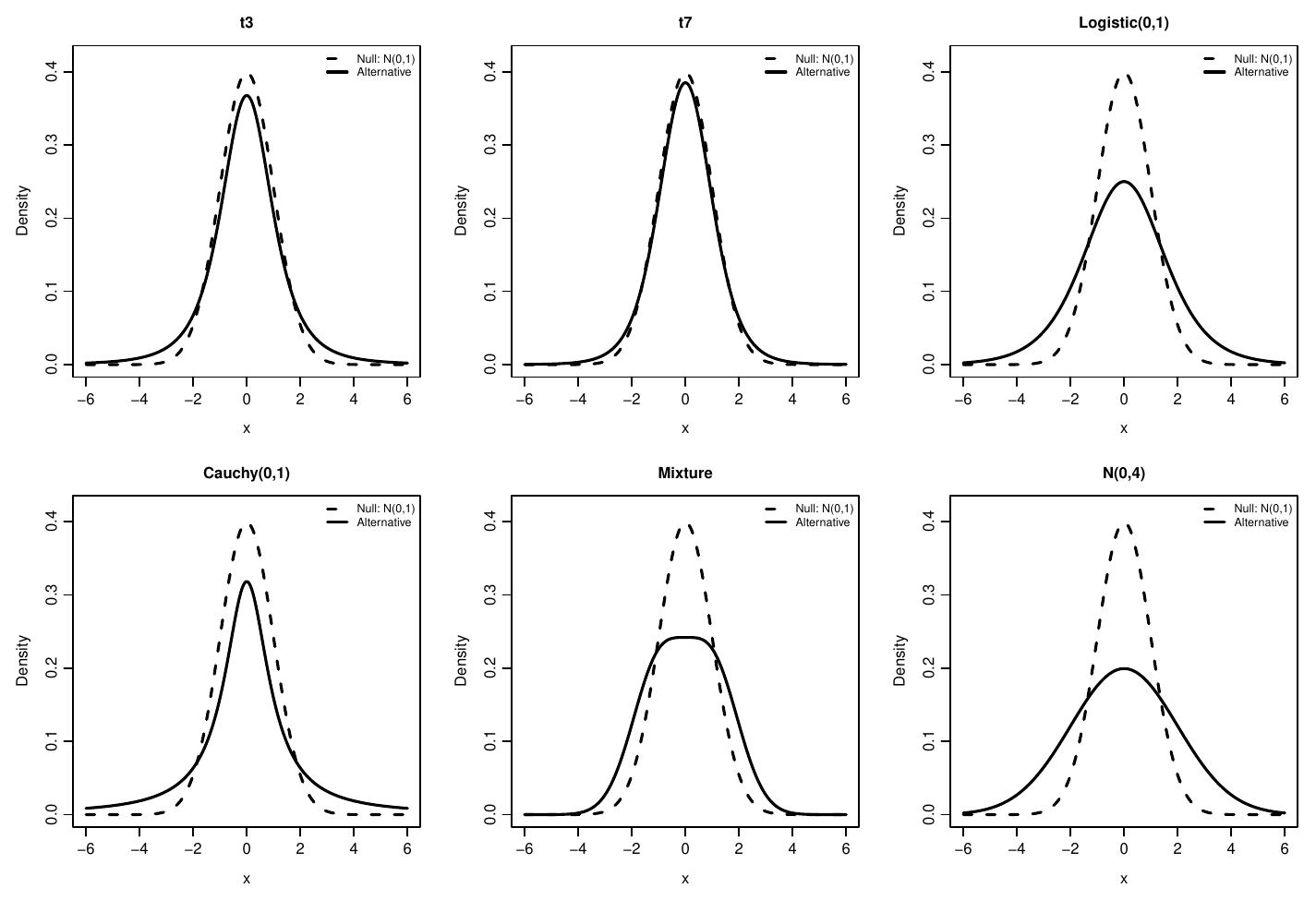}
\caption{\newrev{Overlay of the null $\mathcal{N}(0,1)$ density and the
alternative densities considered in Example~1.}}
\label{fig:ex1densities}
\end{figure}

{
Figure~\ref{fig:ex1densities} overlays the null density with each of the
alternative densities considered in this example. The figure helps illustrate
that the alternatives represent departures of different degrees and forms.
In particular, the $t_7$ density is relatively close to the null
$\mathcal{N}(0,1)$ density, whereas the $t_3$ and Cauchy alternatives exhibit
more pronounced tail departures. The Logistic$(0,1)$ and
$\mathcal{N}(0,4)$ alternatives primarily differ through their spread and
tail behavior, while the mixture represents a departure in shape through
bimodality. Thus, the alternatives provide both relatively subtle and more
pronounced departures from the null.
}

The performance of the test is evaluated using the empirical rejection rate, defined as
\begin{align}
\hat{R} = \frac{1}{M} \sum_{m=1}^M \mathbb{I}\{ \text{reject } H_0 \text{ in the $m$-th replication} \}, \label{rate}    
\end{align}
where $\mathbb{I}(\cdot)$ denotes the indicator function. 

Table~\ref{tab1} reports the empirical rejection rates under the different
data-generating distributions and kernels. {Monte Carlo standard errors
are reported in parentheses to indicate the simulation uncertainty associated
with each estimated rejection rate. With $M=1000$, the standard errors are
small throughout the table, so the broad power patterns are not attributable
to Monte Carlo variation. Under the null, the rejection rates are generally
close to the nominal level $\alpha=0.05$ across the four kernels and sample
sizes, supporting the size calibration of the bootstrap procedure.

The power results are consistent with the visual differences in
Figure~\ref{fig:ex1densities}. The $t_7$ alternative, which is comparatively
close to the null density, produces the most gradual increase in rejection
rates: for the Gaussian kernel, power increases from $0.346$ at $n=30$ to
$0.869$ at $n=200$. In contrast, the more pronounced heavy-tailed $t_3$
alternative is detected much more readily, with corresponding rejection rates
increasing from $0.757$ to $0.999$. The Cauchy, Logistic$(0,1)$, and
$\mathcal{N}(0,4)$ alternatives are detected with very high probability even
at relatively small sample sizes, while the mixture alternative shows a clear
increase from $0.692$ at $n=30$ to essentially one at $n=200$.

The four kernels lead to broadly similar conclusions, although differences are
visible in some of the more difficult small-sample settings. For example,
under $t_7$ with $n=30$, the rejection rates range from $0.292$ for the
uniform kernel to $0.346$ for the Gaussian kernel. We therefore do not regard
the procedure as completely insensitive to kernel choice. Nevertheless, the
overall patterns are similar, and the Gaussian kernel has the additional
practical advantage of avoiding zero leave-one-out density estimates because
of its unbounded support. These considerations support its use as a convenient
default in the comparisons below.}

To further assess the performance of the proposed test, we compare it with a
broader collection of goodness-of-fit procedures. In addition to the
Kolmogorov--Smirnov (KS), Anderson--Darling (AD), and Cram\'er--von Mises
(CvM) tests (\newrev{\hyperref[ref:DS1996]{D'Agostino and Stephens, 1996}}),
{Table~\ref{tabks1} includes Pearson chi-square tests based on
Sturges-rule bins and equal-probability deciles, Zhang's $Z_K$, $Z_C$, and
$Z_A$ likelihood-ratio statistics, and the Shapiro--Wilk (SW) and
Jarque--Bera (JB) normality tests. This expanded comparison includes EDF,
binned, likelihood-ratio-based, and normality-specific benchmarks rather than
restricting the comparison to EDF-based procedures.

The results show that no single procedure dominates in every setting. The
proposed Gaussian-kernel test is particularly competitive for the $t_7$
alternative: at $n=50$, for example, its rejection rate is $0.494$, compared
with $0.432$ for the Sturges Pearson test, $0.445$ for Zhang's $Z_C$, $0.210$
for SW, and $0.281$ for JB. For the heavier-tailed $t_3$ alternative, several
procedures perform similarly well, with the Pearson-Sturges and Zhang tests
sometimes having slightly larger rejection rates than the proposed test. For
the Logistic and Cauchy alternatives, the proposed, Pearson-Sturges, and Zhang
procedures all attain high power as the sample size increases.

The mixture and variance-misspecification alternatives also illustrate why
the comparison depends on the type of departure being considered. The
proposed test has rejection rate $0.692$ for the mixture at $n=30$ and reaches
$0.986$ by $n=100$, while SW and JB have very low rejection rates for this
alternative. Similarly, for $\mathcal{N}(0,4)$ the proposed procedure has
rejection rate one at all reported sample sizes, whereas SW and JB remain near
the nominal level because those procedures test normality rather than the
fully specified variance structure considered here. Consequently, these
comparisons should be interpreted according to the null hypothesis and
alternative targeted by each procedure rather than as evidence that one test
is uniformly superior. The Monte Carlo standard errors in Table~\ref{tabks1}
also provide a direct indication of the simulation uncertainty when comparing
nearby rejection rates.}

\newrev{
We also examined the computational cost of the proposed procedure using
the Gaussian kernel. For each sample size, the elapsed time required to
apply the test to a single data set was recorded using $B=999$ parametric
bootstrap replications. For $n=30,50,100,$ and $200$, the median elapsed
times over five runs were approximately $0.13$, $0.25$, $0.73$, and
$2.42$ seconds, respectively. Under the same bootstrap-calibrated setup,
the corresponding KS, AD, and CvM procedures required approximately
$0.03$--$0.06$ seconds. The additional computational cost of the proposed
procedure is mainly due to the repeated pairwise kernel evaluations
required by the leave-one-out estimator within each bootstrap sample.
Nevertheless, the observed running times remain modest for the sample
sizes considered here. These timing experiments were conducted in
R version 4.6.1 on a laptop equipped with an Intel Core Ultra 9 386H
processor (2.10 GHz) and 32 GB of RAM.
}
\begin{table}[htp]
\centering
\caption{Empirical rejection rates for the KDE-based test in Example 1.
Estimated standard errors of the empirical rejection proportions are given
in parentheses. Results are based on $M=1000$ replications with significance
level $\alpha=0.05$.}
\label{tab1}

\begin{tabular}{llccccc}
\toprule
Data-Generating Distribution & $n$ & $h$ & \multicolumn{4}{c}{Kernel} \\ 
\cmidrule(lr){4-7}
 & & & Gaussian & Epanechnikov & Uniform & Biweight \\
\midrule

\multirow{4}{*}{$\mathcal{N}(0,1)$}
 & 30  & 0.412 & 0.045 (0.007) & 0.041 (0.006) & 0.044 (0.006) & 0.042 (0.006)\\
 & 50  & 0.385 & 0.048 (0.007) & 0.052 (0.007) & 0.055 (0.007) & 0.059 (0.007)\\
 & 100 & 0.343 & 0.048 (0.007) & 0.051 (0.007) & 0.055 (0.007) & 0.051 (0.007)\\
 & 200 & 0.303 & 0.060 (0.008) & 0.064 (0.008) & 0.062 (0.008) & 0.053 (0.007)\\
\midrule

\multirow{4}{*}{$t_3$}
 & 30  & 0.496 & 0.757 (0.014) & 0.763 (0.013) & 0.791 (0.013) & 0.736 (0.014)\\
 & 50  & 0.457 & 0.870 (0.011) & 0.869 (0.011) & 0.907 (0.009) & 0.861 (0.011)\\
 & 100 & 0.404 & 0.993 (0.003) & 0.990 (0.003) & 0.991 (0.003) & 0.983 (0.004)\\
 & 200 & 0.357 & 0.999 (0.001) & 0.999 (0.001) & 1.000 (0.000) & 1.000 (0.000)\\
\midrule

\multirow{4}{*}{$t_7$}
 & 30  & 0.452 & 0.346 (0.015) & 0.309 (0.015) & 0.292 (0.014) & 0.308 (0.015)\\
 & 50  & 0.421 & 0.494 (0.016) & 0.425 (0.016) & 0.408 (0.016) & 0.442 (0.016)\\
 & 100 & 0.373 & 0.684 (0.015) & 0.620 (0.015) & 0.591 (0.016) & 0.601 (0.015)\\
 & 200 & 0.327 & 0.869 (0.011) & 0.824 (0.012) & 0.806 (0.013) & 0.821 (0.012)\\
\midrule

\multirow{4}{*}{Logistic(0,1)}
 & 30  & 0.699 & 0.958 (0.006) & 0.957 (0.006) & 0.970 (0.005) & 0.949 (0.007)\\
 & 50  & 0.645 & 0.995 (0.002) & 0.996 (0.002) & 0.999 (0.001) & 0.994 (0.002)\\
 & 100 & 0.573 & 1.000 (0.000) & 1.000 (0.000) & 1.000 (0.000) & 1.000 (0.000)\\
 & 200 & 0.508 & 1.000 (0.000) & 1.000 (0.000) & 1.000 (0.000) & 1.000 (0.000)\\
\midrule

\multirow{4}{*}{Cauchy(0,1)}
 & 30  & 0.698 & 0.999 (0.001) & 0.999 (0.001) & 0.999 (0.001) & 0.998 (0.001)\\
 & 50  & 0.627 & 1.000 (0.000) & 1.000 (0.000) & 1.000 (0.000) & 1.000 (0.000)\\
 & 100 & 0.541 & 1.000 (0.000) & 1.000 (0.000) & 1.000 (0.000) & 1.000 (0.000)\\
 & 200 & 0.467 & 1.000 (0.000) & 1.000 (0.000) & 1.000 (0.000) & 1.000 (0.000)\\
\midrule

\multirow{4}{*}{Mixture}
 & 30  & 0.616 & 0.692 (0.015) & 0.696 (0.015) & 0.794 (0.013) & 0.703 (0.014)\\
 & 50  & 0.569 & 0.877 (0.010) & 0.884 (0.010) & 0.936 (0.008) & 0.880 (0.010)\\
 & 100 & 0.500 & 0.986 (0.004) & 0.992 (0.003) & 0.999 (0.001) & 0.987 (0.004)\\
 & 200 & 0.440 & 1.000 (0.000) & 1.000 (0.000) & 1.000 (0.000) & 1.000 (0.000)\\
\midrule

\multirow{4}{*}{$\mathcal{N}(0,4)$}
 & 30  & 1.649 & 1.000 (0.000) & 1.000 (0.000) & 1.000 (0.000) & 1.000 (0.000)\\
 & 50  & 1.539 & 1.000 (0.000) & 1.000 (0.000) & 1.000 (0.000) & 1.000 (0.000)\\
 & 100 & 1.373 & 1.000 (0.000) & 1.000 (0.000) & 1.000 (0.000) & 1.000 (0.000)\\
 & 200 & 1.210 & 1.000 (0.000) & 1.000 (0.000) & 1.000 (0.000) & 1.000 (0.000)\\
\bottomrule
\end{tabular}
\end{table}

\begin{sidewaystable}[p]
\centering

\caption{Empirical rejection rates for the KDE-based test and competing
goodness-of-fit tests in Example 1. Monte Carlo standard errors are given
in parentheses ($M=1000$, $\alpha=0.05$).}
\label{tabks1}

\vspace{2mm}

\resizebox{0.95\textheight}{!}{%
\begin{tabular}{llccccccccccc}
\toprule
& & \multicolumn{1}{c}{Proposed}
& \multicolumn{3}{c}{EDF}
& \multicolumn{2}{c}{Pearson}
& \multicolumn{3}{c}{Zhang}
& \multicolumn{2}{c}{Normality} \\
\cmidrule(lr){3-3}
\cmidrule(lr){4-6}
\cmidrule(lr){7-8}
\cmidrule(lr){9-11}
\cmidrule(lr){12-13}
Distribution & $n$ & KDE & KS & AD & CvM
& $\chi^2_S$ & $\chi^2_D$
& $Z_K$ & $Z_C$ & $Z_A$
& SW & JB \\
\midrule

\multirow{4}{*}{$\mathcal{N}(0,1)$}
& 30  & 0.045 (0.007) & 0.058 (0.007) & 0.053 (0.007) & 0.054 (0.007)
& 0.046 (0.007) & 0.048 (0.007) & 0.047 (0.007) & 0.051 (0.007) & 0.049 (0.007) & 0.053 (0.007) & 0.050 (0.007) \\
& 50  & 0.048 (0.007) & 0.046 (0.007) & 0.046 (0.007) & 0.050 (0.007)
& 0.042 (0.006) & 0.043 (0.006) & 0.042 (0.006) & 0.050 (0.007) & 0.051 (0.007) & 0.058 (0.007) & 0.059 (0.007) \\
& 100 & 0.048 (0.007) & 0.054 (0.007) & 0.064 (0.008) & 0.060 (0.008)
& 0.048 (0.007) & 0.052 (0.007) & 0.049 (0.007) & 0.053 (0.007) & 0.060 (0.008) & 0.058 (0.007) & 0.054 (0.007) \\
& 200 & 0.060 (0.008) & 0.066 (0.008) & 0.080 (0.009) & 0.085 (0.009)
& 0.056 (0.007) & 0.047 (0.007) & 0.046 (0.007) & 0.051 (0.007) & 0.053 (0.007) & 0.043 (0.006) & 0.047 (0.007) \\
\midrule

\multirow{4}{*}{$t_3$}
& 30  & 0.757 (0.014) & 0.060 (0.008) & 0.401 (0.015) & 0.064 (0.008)
& 0.763 (0.013) & 0.135 (0.011) & 0.764 (0.013) & 0.784 (0.013) & 0.744 (0.014) & 0.454 (0.016) & 0.508 (0.016) \\
& 50  & 0.870 (0.011) & 0.088 (0.009) & 0.549 (0.016) & 0.098 (0.009)
& 0.910 (0.009) & 0.186 (0.012) & 0.892 (0.010) & 0.919 (0.009) & 0.881 (0.010) & 0.635 (0.015) & 0.709 (0.014) \\
& 100 & 0.993 (0.003) & 0.127 (0.011) & 0.766 (0.013) & 0.155 (0.011)
& 0.988 (0.003) & 0.308 (0.015) & 0.985 (0.004) & 0.990 (0.003) & 0.986 (0.004) & 0.872 (0.011) & 0.898 (0.010) \\
& 200 & 0.999 (0.001) & 0.238 (0.013) & 0.970 (0.005) & 0.305 (0.015)
& 1.000 (0.000) & 0.590 (0.016) & 1.000 (0.000) & 1.000 (0.000) & 1.000 (0.000) & 0.995 (0.002) & 0.996 (0.002) \\
\midrule

\multirow{4}{*}{$t_7$}
& 30  & 0.346 (0.015) & 0.069 (0.008) & 0.117 (0.010) & 0.067 (0.008)
& 0.303 (0.015) & 0.036 (0.006) & 0.279 (0.014) & 0.336 (0.015) & 0.270 (0.014) & 0.164 (0.012) & 0.220 (0.013) \\
& 50  & 0.494 (0.016) & 0.076 (0.008) & 0.137 (0.011) & 0.071 (0.008)
& 0.432 (0.016) & 0.064 (0.008) & 0.368 (0.015) & 0.445 (0.016) & 0.347 (0.015) & 0.210 (0.013) & 0.281 (0.014) \\
& 100 & 0.684 (0.015) & 0.092 (0.009) & 0.225 (0.013) & 0.108 (0.010)
& 0.634 (0.015) & 0.074 (0.008) & 0.569 (0.016) & 0.638 (0.015) & 0.552 (0.016) & 0.365 (0.015) & 0.437 (0.016) \\
& 200 & 0.869 (0.011) & 0.112 (0.010) & 0.290 (0.014) & 0.139 (0.011)
& 0.850 (0.011) & 0.122 (0.010) & 0.811 (0.012) & 0.872 (0.011) & 0.823 (0.012) & 0.577 (0.016) & 0.670 (0.015) \\
\midrule

\multirow{4}{*}{Logistic$(0,1)$}
& 30  & 0.958 (0.006) & 0.345 (0.015) & 0.849 (0.011) & 0.368 (0.015)
& 0.959 (0.006) & 0.481 (0.016) & 0.966 (0.006) & 0.972 (0.005) & 0.966 (0.006) & 0.150 (0.011) & 0.196 (0.013) \\
& 50  & 0.995 (0.002) & 0.530 (0.016) & 0.964 (0.006) & 0.578 (0.016)
& 0.999 (0.001) & 0.760 (0.014) & 0.996 (0.002) & 0.999 (0.001) & 0.999 (0.001) & 0.183 (0.012) & 0.242 (0.014) \\
& 100 & 1.000 (0.000) & 0.884 (0.010) & 0.999 (0.001) & 0.929 (0.008)
& 1.000 (0.000) & 0.988 (0.003) & 1.000 (0.000) & 1.000 (0.000) & 1.000 (0.000) & 0.300 (0.014) & 0.363 (0.015) \\
& 200 & 1.000 (0.000) & 1.000 (0.000) & 1.000 (0.000) & 1.000 (0.000)
& 1.000 (0.000) & 1.000 (0.000) & 1.000 (0.000) & 1.000 (0.000) & 1.000 (0.000) & 0.509 (0.016) & 0.616 (0.015) \\
\midrule

\multirow{4}{*}{Cauchy$(0,1)$}
& 30  & 0.999 (0.001) & 0.286 (0.014) & 0.987 (0.004) & 0.276 (0.014)
& 0.998 (0.001) & 0.800 (0.013) & 0.998 (0.001) & 0.999 (0.001) & 0.999 (0.001) & 0.958 (0.006) & 0.955 (0.007) \\
& 50  & 1.000 (0.000) & 0.503 (0.016) & 1.000 (0.000) & 0.473 (0.016)
& 1.000 (0.000) & 0.946 (0.007) & 1.000 (0.000) & 1.000 (0.000) & 1.000 (0.000) & 0.999 (0.001) & 0.995 (0.002) \\
& 100 & 1.000 (0.000) & 0.854 (0.011) & 1.000 (0.000) & 0.856 (0.011)
& 1.000 (0.000) & 0.989 (0.003) & 1.000 (0.000) & 1.000 (0.000) & 1.000 (0.000) & 1.000 (0.000) & 1.000 (0.000) \\
& 200 & 1.000 (0.000) & 1.000 (0.000) & 1.000 (0.000) & 0.997 (0.002)
& 1.000 (0.000) & 1.000 (0.000) & 1.000 (0.000) & 1.000 (0.000) & 1.000 (0.000) & 1.000 (0.000) & 1.000 (0.000) \\
\midrule

\multirow{4}{*}{Mixture}
& 30  & 0.692 (0.015) & 0.292 (0.014) & 0.593 (0.016) & 0.296 (0.014)
& 0.690 (0.015) & 0.345 (0.015) & 0.695 (0.015) & 0.662 (0.015) & 0.662 (0.015) & 0.045 (0.007) & 0.014 (0.004) \\
& 50  & 0.877 (0.010) & 0.448 (0.016) & 0.803 (0.013) & 0.473 (0.016)
& 0.858 (0.011) & 0.610 (0.015) & 0.874 (0.010) & 0.853 (0.011) & 0.841 (0.012) & 0.066 (0.008) & 0.010 (0.003) \\
& 100 & 0.986 (0.004) & 0.735 (0.014) & 0.980 (0.004) & 0.819 (0.012)
& 0.990 (0.003) & 0.950 (0.007) & 0.995 (0.002) & 0.994 (0.002) & 0.994 (0.002) & 0.111 (0.010) & 0.008 (0.003) \\
& 200 & 1.000 (0.000) & 0.990 (0.003) & 1.000 (0.000) & 0.997 (0.002)
& 1.000 (0.000) & 1.000 (0.000) & 1.000 (0.000) & 1.000 (0.000) & 1.000 (0.000) & 0.235 (0.013) & 0.075 (0.008) \\
\midrule

\multirow{4}{*}{$\mathcal{N}(0,4)$}
& 30  & 1.000 (0.000) & 0.607 (0.015) & 0.990 (0.003) & 0.658 (0.015)
& 0.992 (0.003) & 0.791 (0.013) & 0.995 (0.002) & 0.995 (0.002) & 0.996 (0.002) & 0.046 (0.007) & 0.049 (0.007) \\
& 50  & 1.000 (0.000) & 0.840 (0.012) & 0.998 (0.001) & 0.892 (0.010)
& 1.000 (0.000) & 0.975 (0.005) & 1.000 (0.000) & 1.000 (0.000) & 1.000 (0.000) & 0.047 (0.007) & 0.055 (0.007) \\
& 100 & 1.000 (0.000) & 0.997 (0.002) & 1.000 (0.000) & 1.000 (0.000)
& 1.000 (0.000) & 1.000 (0.000) & 1.000 (0.000) & 1.000 (0.000) & 1.000 (0.000) & 0.051 (0.007) & 0.055 (0.007) \\
& 200 & 1.000 (0.000) & 1.000 (0.000) & 1.000 (0.000) & 1.000 (0.000)
& 1.000 (0.000) & 1.000 (0.000) & 1.000 (0.000) & 1.000 (0.000) & 1.000 (0.000) & 0.053 (0.007) & 0.067 (0.008) \\
\bottomrule
\end{tabular}%
}

\vspace{2mm}

\begin{minipage}{0.78\textheight}
\scriptsize
\textit{Note:} Values in parentheses are Monte Carlo standard errors.
$\chi^2_S$ and $\chi^2_D$ denote the Pearson chi-square tests using
Sturges-rule bins and deciles, respectively; $Z_K$, $Z_C$, and $Z_A$
are Zhang statistics; SW = Shapiro--Wilk; JB = Jarque--Bera.
\end{minipage}

\end{sidewaystable}

\textbf{Example 2.} \textit{Testing for Exponentiality (Boundary Case):} We next test the null hypothesis $H_0\!: f = f_0$, where $f_0$ is the density of an exponential distribution with mean $\lambda$ estimated by $\hat{\lambda} = \bar{x}$. To assess sensitivity, we generate data from Gamma(2,1), Weibull(1.5,1), and Lognormal(0,1). Since the support is $[0, \infty)$, we employ the reflection method at the boundary $a = 0$ to mitigate bias in the kernel estimator.

Table~\ref{tab2} presents the empirical rejection rates as defined in (\ref{rate}). Under the null, the rejection rates align closely with the 5\% nominal level, confirming size accuracy. Against alternatives, the test displays increasing power as $n$ grows, particularly for the Gamma and Weibull cases, where near-perfect power is achieved for $n \geq 100$. For the more challenging Lognormal alternative, power rises steadily with $n$, attaining almost one for $n = 200$. This example underscores the utility of boundary correction in maintaining accuracy and the flexibility of the test in detecting departures from exponentiality. We also compare the performance of the proposed test using the Gaussian kernel with that of the Kolmogorov--Smirnov, Anderson--Darling, and Cram\'er--von Mises
tests. \newrev{The results are reported in Table~\ref{tabks2} and show that the relative performance varies across alternatives and sample sizes.}

\begin{table}[htp]
\centering
\caption{Empirical rejection rates for the KDE-based test in Example 2. Results are based on $M = 1000$ replications, with significance level $\alpha = 0.05$.}
\label{tab2}
\begin{tabular}{lllcccc}
\toprule
Data-Generating Distribution& $n$ & $h$ & \multicolumn{4}{c}{Kernel} \\ 
\cmidrule(lr){4-7}
             &     &     & Gaussian & Epanechnikov & Uniform & Biweight \\
\midrule
\multirow{4}{*}{Exp(1)}
 & 30  &0.351 &0.024 & 0.020 &0.022 &0.018\\
 & 50  &0.326 &0.030 &0.038 &0.036 &0.044\\
 & 100 &0.290 &0.042 &0.043 &0.039 &0.039 \\
 & 200 &0.253 &0.050 &0.046 &0.043 &0.043\\
\midrule
\multirow{4}{*}{Gamma(2,1)}
 & 30  &0.544 &0.178 &0.244 & 0.515 &0.376 \\
 & 50  &0.503 &0.438 &0.558 &0.860 &0.693 \\
 & 100 &0.450 &0.923 &0.960 & 1 & 0.984\\
 & 200 &0.397 &1 &1 &1 &1\\
\midrule
\multirow{4}{*}{Weibull(1.5,1)}
 & 30  &0.249 &0.139 &0.195 &0.464 &0.304\\
 & 50  &0.230 &0.354 &0.468 &0.800 &0.631 \\
 & 100 &0.205 &0.831 &0.911 &0.989 &0.955 \\
 & 200 &0.183 &0.998 &0.999 &1 &0.999\\
\midrule
\multirow{4}{*}{Lognormal(0,1)}
 & 30  &0.477 &0.304 &0.321 &0.304 & 0.322\\
 & 50  &0.432 &0.517 &0.540 &0.509 &0.524 \\
 & 100 &0.382 &0.871 &0.882 &0.844 &0.856 \\
 & 200 &0.338 &0.999 &0.998 &0.999 &0.998 \\
\bottomrule
\end{tabular}
\end{table}

\begin{table}[!htp]
\centering
\caption{Comparison between the KDE-based test with Gaussian kernel and standard goodness-of-fit tests. The simulation is based on $M = 1000$ replications with significance level $\alpha = 0.05$.}
\label{tabks2}
\begin{tabular}{llcccc}
\toprule
Data-Generating Distribution & $n$ & KDE & KS & AD & CvM \\
\midrule
\multirow{4}{*}{$\mathrm{Exp}(1)$}
 & 30  & 0.024 & 0.049 & 0.048 & 0.050\\
 & 50  & 0.030 & 0.045 & 0.052 & 0.054\\
 & 100 & 0.042 & 0.043 & 0.049 & 0.047\\
 & 200 & 0.050 & 0.044 & 0.058 & 0.057\\
\midrule
\multirow{4}{*}{$\mathrm{Gamma}(2,1)$}
 & 30  & 0.178 & 0.999 & 0.999 & 1.000\\
 & 50  & 0.438 & 1.000 & 1.000 & 1.000\\
 & 100 & 0.923 & 1.000 & 1.000 & 1.000\\
 & 200 & 1.000 & 1.000 & 1.000 & 1.000\\
\midrule
\multirow{4}{*}{$\mathrm{Weibull}(1.5,1)$}
 & 30  & 0.139 & 0.130 & 0.128 & 0.114\\
 & 50  & 0.354 & 0.258 & 0.362 & 0.279\\
 & 100 & 0.831 & 0.585 & 0.880 & 0.735\\
 & 200 & 0.998 & 0.967 & 1.000 & 0.992\\
\midrule
\multirow{4}{*}{$\mathrm{Lognormal}(0,1)$}
 & 30  & 0.304 & 0.418 & 0.676 & 0.545\\
 & 50  & 0.517 & 0.680 & 0.898 & 0.801\\
 & 100 & 0.871 & 0.947 & 0.993 & 0.983\\
 & 200 & 0.999 & 1.000 & 1.000 & 1.000\\
\bottomrule
\end{tabular}
\end{table}

\subsection{Multivariate Examples}
\textbf{Example 3.} \textit{Testing Multivariate Normality.}  
We now turn to the multivariate case with dimension $d = 2$ and test the null hypothesis
$$
H_0: f = f_0, \quad \text{where} \quad f_0 = \phi_2(\cdot; \boldsymbol{0}, I_2),
$$
the bivariate normal density with mean vector $\boldsymbol{0}$ and identity covariance matrix $I_2$. For each simulated sample, the KDE-based test is applied using two multivariate kernels: Gaussian and Epanechnikov. The estimator $\hat{f}_n(\mathbf{x})$ is constructed as in~\eqref{multi_1}, with a diagonal bandwidth matrix having entries $h_j$ as defined in~\eqref{h2}. 

To examine the sensitivity of the proposed test, we generate samples from the following alternative distributions:
\begin{itemize}
    \item Bivariate $t_3$ distribution (heavy-tailed),
    \item Contaminated normal: $0.9\, \mathcal{N}(\boldsymbol{0}, I_2) + 0.1\, \mathcal{N}(\boldsymbol{3}, I_2)$ (bimodal structure),
    \item Uniform distribution on $[0,1]^2$ (non-elliptical support).
\end{itemize}

Table~\ref{tab3} reports the empirical rejection rates based on $M = 1000$ replications, with significance level $\alpha = 0.05$. 
\newrev{
The proposed KDE-based test exhibits high rejection rates against the alternatives considered, with rejection probabilities increasing rapidly with sample size for the heavy-tailed and contaminated-normal alternatives and reaching one for the uniform alternative.
}
To further evaluate its performance, we compare it with several multivariate normality goodness-of-fit tests, including Mardia’s skewness and kurtosis tests (\newrev{\hyperref[ref:M1970]{Mardia, 1970}}; \newrev{\hyperref[ref:M1974]{Mardia, 1974}}), Henze–Zirkler’s test (\newrev{\hyperref[ref:HZ1990]{Henze and Zirkler, 1990}}), and Royston’s test (\newrev{\hyperref[ref:R1983]{Royston, 1983}}). These tests are implemented using the \texttt{MVN} package in \textsf{R} (\newrev{\hyperref[ref:KGZ2014]{Korkmaz et al., 2014}}).

\newrev{
We note that these procedures assess general multivariate normality,
whereas the null hypothesis in the present simulation is the fully
specified $N_2(\boldsymbol{0},I_2)$ distribution. Therefore, the
comparisons in Table~\ref{tabks3} should be interpreted as comparisons
with commonly used multivariate normality benchmarks rather than as
exact like-for-like power comparisons under identical null hypotheses.
The results show that the proposed procedure has high rejection rates
against the alternatives considered, particularly for the contaminated
normal and uniform alternatives, but we do not interpret these results
as establishing uniform superiority over the competing procedures.
}

\begin{table}[htp]
\centering
\caption{Empirical rejection rates for the KDE-based test in Example 3. Results are based on $M = 1000$ replications, with significance level $\alpha = 0.05$.}
\label{tab3}
\begin{tabular}{lllcc}
\toprule
Data-Generating Distributions & $h=(h_1,h_2)$ & $n$ & \multicolumn{2}{c}{Kernel} \\ 
\cmidrule(lr){4-5}
             &     &     & Gaussian & Epanechnikov \\
%Scenario & $h=(h_1,h_2)$ & $n$ & Gaussian & Epanechnikov \\
\midrule
\multirow{4}{*}{$\mathcal{N}(\boldsymbol{0}, I_2)$}
&(0.564, 0.561) & 30   & 0.054 & 0.048 \\
&(0.520, 0.517) & 50   & 0.067 & 0.057 \\
&(0.461, 0.461) & 100  & 0.059 & 0.059 \\
&(0.414, 0.413) & 200  & 0.050 & 0.043 \\
\midrule
\multirow{4}{*}{Bivariate $t_3$}
&(0.914, 0.916) & 30   & 0.910 & 0.870 \\
&(0.861, 0.875) & 50   & 0.980 & 0.970 \\
&(0.770, 0.778) & 100  & 1.000 & 1.000 \\
&(0.699, 0.693) & 200  & 1.000 & 1.000 \\
\midrule
\multirow{4}{*}{Contaminated Normal}
&(0.763, 0.759) & 30   & 0.864 & 0.812 \\
&(0.696, 0.696) & 50   & 0.964 & 0.934 \\
&(0.622, 0.623) & 100  & 0.999 & 0.995 \\
&(0.556, 0.557) & 200  & 1.000 & 1.000 \\
\midrule
\multirow{4}{*}{Uniform on $[0,1]^2$}
&(0.164, 0.163) & 30   & 1.000 & 1.000 \\
&(0.150, 0.150) & 50   & 1.000 & 1.000 \\
&(0.134, 0.134) & 100  & 1.000 & 1.000 \\
&(0.119, 0.119) & 200  & 1.000 & 1.000 \\
\bottomrule
\end{tabular}
\end{table}

\begin{table}[!htp]
\centering
\caption{Comparison between the KDE-based test with Gaussian kernel and standard goodness-of-fit tests. 
The simulation is based on $M = 1000$ replications with a significance level $\alpha = 0.05$.
\newrev{The competing procedures are included as commonly used multivariate normality benchmarks and do not all test the same fully specified null hypothesis as the KDE-based procedure.}}
\label{tabks3}
\begin{tabular}{lcccccc}
\toprule
\makecell{Data-Generating\\Distribution} & $n$ & KDE & Mardia skew. & Mardia kurt. & Henze--Zirkler & Royston \\
\midrule
\multirow{4}{*}{$\mathcal{N}(\boldsymbol{0}, I_2)$}
 & 30  & 0.054 & 0.030 & 0.013 & 0.057 & 0.064 \\
 & 50  & 0.067 & 0.039 & 0.016 & 0.048 & 0.057 \\
 & 100 & 0.059 & 0.043 & 0.019 & 0.053 & 0.053 \\
 & 200 & 0.050 & 0.054 & 0.027 & 0.042 & 0.051 \\
\midrule
\multirow{4}{*}{Bivariate $t_3$}
 & 30  & 0.910 & 0.313 & 0.287 & 0.425 & 0.498 \\
 & 50  & 0.980 & 0.429 & 0.453 & 0.562 & 0.672 \\
 & 100 & 1.000 & 0.485 & 0.800 & 0.795 & 0.900 \\
 & 200 & 1.000 & 0.539 & 1.000 & 1.000 & 1.000 \\
\midrule
\multirow{4}{*}{Contaminated Normal}
 & 30  & 0.864 & 0.463 & 0.123 & 0.468 & 0.494 \\
 & 50  & 0.964 & 0.715 & 0.227 & 0.654 & 0.665 \\
 & 100 & 0.999 & 0.910 & 0.414 & 0.807 & 0.836 \\
 & 200 & 1.000 & 1.000 & 0.698 & 0.923 & 1.000 \\
\midrule
\multirow{4}{*}{Uniform on $[0,1]^2$}
 & 30  & 1.000 & 0.000 & 0.073 & 0.347 & 0.578 \\
 & 50  & 1.000 & 0.000 & 0.612 & 0.659 & 0.939 \\
 & 100 & 1.000 & 0.000 & 0.999 & 0.970 & 1.000 \\
 & 200 & 1.000 & 0.001 & 1.000 & 1.000 & 1.000 \\
\bottomrule
\end{tabular}
\end{table}

\subsection{Real Data Applications}

In this section, we demonstrate the practical performance of the proposed KDE-based goodness-of-fit test using a selection of real datasets. Both univariate and multivariate applications are considered.

\medskip

\noindent\textbf{Example 4.} \textit{Testing a Laplace Model (Yarn Breaking Strength Data).}  
We analyze the classical dataset of breaking strength measurements for 100 yarns, originally reported by \newrev{\hyperref[ref:D1974]{Duncan (1974)}}. The observations are univariate and continuous, listed below:
$$
\begin{array}{rrrrrrrrrrrrrrr}
66  & 92  & 99  & 94  & 117 & 137 & 85  & 105 & 132 & 91  & 95  & 103 & 111 & 84  & 89 \\
96  & 107 & 96  & 102 & 100 & 85  & 97  & 100 & 101 & 89  & 100 & 98  & 98  & 79  & 105 \\
97  & 97  & 91  & 104 & 104 & 97  & 97  & 137 & 114 & 101 & 138 & 80  & 111 & 102 & 103 \\
104 & 98  & 98  & 111 & 104 & 99  & 94  & 86  & 106 & 102 & 100 & 78  & 84  & 91  & 98  \\
96  & 92  & 95  & 99  & 93  & 86  & 111 & 92  & 101 & 104 & 104 & 102 & 102 & 132 & 97  \\
87  & 110 & 94  & 98  & 99  & 95  & 99  & 102 & 62  & 96  & 102 & 109 & 92  & 88  & 101 \\
88  & 100 & 122 & 104 & 91  & 96  & 115 & 107 & 103 & 98  &     &     &     &     &     
\end{array}
$$

We test the null hypothesis $H_0\!: f = f_0$, where $f_0$ is the density of the Laplace$(\mu, \sigma)$ distribution,  
$$
f_0(x) = \frac{1}{2\sigma} \exp\!\left(-\frac{|x - \mu|}{\sigma}\right), \quad x \in \mathbb{R},
$$
with location parameter $\mu \in \mathbb{R}$ and scale parameter $\sigma > 0$. The maximum likelihood estimates are $\hat{\mu} = 99$ and $\hat{\sigma} = 8.33$.  

The KDE-based test was implemented using a Gaussian kernel with bandwidth selected according to~\eqref{h1_rev}, yielding $h = 2.741$. The test statistic was $\log T_n = 9.655$, and with $B = 999$ bootstrap replications the resulting $p$-value was $p_{\text{boot}} = 0.210$, providing no evidence against the Laplace model. This outcome supports earlier findings and confirms the adequacy of the parametric fit. For comparison, we also applied several classical goodness-of-fit tests, including the Kolmogorov--Smirnov, Anderson--Darling, and Cram\'er--von Mises. The $p$-values are 0.574, 0.764, 0.872, respectively. All tests, consistent with the KDE-based method, fail to reject the null hypothesis at the 5\% significance level, reinforcing the conclusion that the Laplace distribution provides a satisfactory model for the yarn strength data.  

\medskip
\noindent\textbf{Example 5.} \textit{Testing a Gamma Model (River Flow Data).}  
Next, we analyze river flow measurements (in millions of cubic feet per second) from the Susquehanna River at Harrisburg, Pennsylvania, recorded during 1980–1984. The dataset consists of twenty observations:
$$
\begin{array}{cccccccccc}
0.654,& 0.613,& 0.315,& 0.449,& 0.297,& 0.402,& 0.379,& 0.423,& 0.379,& 0.324,\\
0.269,& 0.740,& 0.418,& 0.412,& 0.494,& 0.416,& 0.338,& 0.392,& 0.484,& 0.265.
\end{array}
$$

We assess whether a three-parameter Gamma distribution provides an adequate fit. That is, we test $H_0\!: f=f_0$, where $f_0$ denotes the Gamma$(\alpha, \beta, \gamma)$ density given by
$$
f(x; \alpha, \beta, \gamma) \;=\; \frac{\beta^{-\alpha} (x - \gamma)^{\alpha - 1} e^{-(x - \gamma)/\beta}}{\Gamma(\alpha)}, 
\qquad x > \gamma,
$$
with $\alpha > 0$, $\beta > 0$, and $\gamma > 0$ the shape, scale, and location parameters, respectively.
Maximum likelihood estimates yield $\hat{\alpha} = 1.705$, $\hat{\beta} = 0.096$, and $\hat{\gamma} = 0.260$. 
Since the data are supported on $[0,\infty)$, boundary bias is corrected using the reflection method with $a=0$. 
The bandwidth is chosen according to~\eqref{h1_rev}, giving $h = 0.045$.

The KDE-based test is implemented with several kernels. Table~\ref{gamma-kde} reports the bootstrap $p$-values (based on $B = 999$ replicates) under $H_0$ with the fixed MLEs. 
All kernels yield non-significant results at level $0.05$, providing no evidence against the Gamma model.

\begin{table}[H]
\centering
\caption{KDE-based test for Gamma$(\hat{\alpha},\hat{\beta},\hat{\gamma})$ on river flow data: 
bootstrap $p$-values (bandwidth $h=0.045$, $B = 999$).}
\label{gamma-kde}
\begin{tabular}{lc}
\toprule
Kernel & $p$-value \\
\midrule
Gaussian      & 0.534 \\
Epanechnikov  & 0.570 \\
Uniform       & 0.338 \\
Biweight      & 0.628 \\
\bottomrule
\end{tabular}
\end{table}

For comparison, we also apply classical goodness-of-fit tests 
(Kolmogorov--Smirnov, Anderson--Darling, and Cram\'er--von Mises). The corresponding $p$-values 0.714, 0.750, and 0.874.  Consistent with the KDE-based results, all tests fail to reject $H_0$ at the $5\%$ level, reinforcing the conclusion that the three-parameter Gamma model provides an adequate fit.

\medskip

\newrev{
\noindent\textbf{Example 6.} \textit{Testing Four-Dimensional Multivariate Normality (Iris Setosa Data).}  
We now analyze the joint distribution of the variables
\texttt{Sepal.Length}, \texttt{Sepal.Width},
\texttt{Petal.Length}, and \texttt{Petal.Width}
from the \texttt{setosa} species of the \texttt{iris} dataset,
consisting of $n=50$ observations. This dataset has also been used
in the literature to assess multivariate normality; in particular,
\hyperref[ref:KGZ2014]{Korkmaz et al. (2014)} considered the same
four-dimensional \textit{Iris setosa} data using several multivariate
normality tests. The null hypothesis is that the data arise from a
four-dimensional multivariate normal distribution
$N_4(\boldsymbol{\mu},\Sigma)$:
$$
H_0: f = f_0, \quad
f_0(\mathbf{x};\boldsymbol{\mu},\Sigma)
=
\frac{1}{(2\pi)^2|\Sigma|^{1/2}}
\exp\!\left(
-\tfrac{1}{2}
(\mathbf{x}-\boldsymbol{\mu})^\top
\Sigma^{-1}
(\mathbf{x}-\boldsymbol{\mu})
\right),
$$
where $\boldsymbol{\mu}\in\mathbb{R}^4$ and $\Sigma$ is a
positive-definite covariance matrix.

Maximum likelihood estimates yield
$$
\hat{\boldsymbol{\mu}}
=
(5.006,\;3.428,\;1.462,\;0.246)^\top
$$
and
$$
\hat{\Sigma}
=
\begin{pmatrix}
0.1218 & 0.0972 & 0.0160 & 0.0101 \\
0.0972 & 0.1408 & 0.0115 & 0.0091 \\
0.0160 & 0.0115 & 0.0296 & 0.0059 \\
0.0101 & 0.0091 & 0.0059 & 0.0109
\end{pmatrix}.
$$

The KDE-based test is implemented using the Gaussian and
Epanechnikov product kernels, with bandwidths selected according to
rule~\eqref{h2}. The resulting bandwidth vector is
$$
(0.2055,\;0.2210,\;0.1012,\;0.0614).
$$

For the Gaussian-kernel implementation, the bootstrap $p$-value is
computed using $B=999$ replicates. In each bootstrap sample, the mean
vector, covariance matrix, bandwidths, and leave-one-out KDE are
recomputed. The Epanechnikov implementation is also evaluated for the
observed data; because zero leave-one-out density estimates occur, the
corresponding statistic is undefined and bootstrap calibration is not
performed.

For the Gaussian kernel, all leave-one-out density estimates are
positive and the resulting bootstrap $p$-value is $0.623$. In
contrast, the Epanechnikov kernel produces zero leave-one-out density
estimates for 24 of the 50 observations. Consequently, the
corresponding log-statistic is undefined and a bootstrap $p$-value
is not reported. This illustrates a practical limitation of
compact-support kernels in the multivariate setting and supports the
use of the Gaussian kernel as a safer default choice.

As in Example 3, we compare the results with several standard
multivariate normality tests, including Mardia's skewness and kurtosis
tests
(\hyperref[ref:M1970]{Mardia, 1970};
\hyperref[ref:M1974]{Mardia, 1974}),
the Henze--Zirkler test
(\hyperref[ref:HZ1990]{Henze and Zirkler, 1990}),
and Royston's test
(\hyperref[ref:R1983]{Royston, 1983}).
In response to the reviewer's suggestion, we also include the
Doornik--Hansen test
(\hyperref[ref:DH2008]{Doornik and Hansen, 2008}),
an energy-distance test
(\hyperref[ref:SR2005]{Sz\'ekely and Rizzo, 2005}),
and an MMD-based test
(\hyperref[ref:GBRS2012]{Gretton et al., 2012}).
The results are summarized in Table~\ref{tab5}.

\begin{table}[htp]
\centering
\caption{$p$-values of multivariate normality tests for the
four-dimensional \textit{Iris setosa} data.}
\label{tab5}
\begin{tabular}{lc}
\toprule
Test & $p$-value \\
\midrule
KDE-based (Gaussian kernel)      & 0.623 \\
KDE-based (Epanechnikov kernel)  & Undefined$^{\dagger}$ \\
Mardia skewness                  & 0.177 \\
Mardia kurtosis                  & 0.195 \\
Henze--Zirkler                   & 0.04995 \\
Royston                          & $<0.001$ \\
Doornik--Hansen                  & $<0.001$ \\
Energy                           & 0.035 \\
MMD                              & 0.049 \\
\bottomrule
\end{tabular}

\medskip
\footnotesize
$^{\dagger}$The Epanechnikov product kernel gives zero leave-one-out
density estimates for 24 observations, so the corresponding
log-statistic is undefined.
\end{table}

The results show clear differences among the multivariate normality
tests. The KDE-based test with the Gaussian kernel does not reject
$H_0$ at the $5\%$ level, with a bootstrap $p$-value of $0.623$.
Similarly, Mardia's skewness and kurtosis tests do not reject
multivariate normality. In contrast, Royston's test, the
Doornik--Hansen test, the energy test, and the MMD test reject
$H_0$ at the $5\%$ level, while the Henze--Zirkler $p$-value is
$0.04995$, which is just below the $5\%$ significance level.

These differences are consistent with previously reported analyses
of the same dataset. In particular,
\hyperref[ref:KGZ2014]{Korkmaz et al. (2014)} reported
$p$-values of $0.1772$ and $0.1953$ for Mardia's skewness and
kurtosis tests, respectively, $0.04995$ for the Henze--Zirkler
test, and a very small $p$-value for Royston's test. Thus, the
different conclusions obtained from the competing procedures are not
specific to the present implementation, but have also been observed
in previous analyses of the \textit{Iris setosa} data.

This example illustrates the application of the proposed procedure
beyond the two-dimensional setting considered in Example 3. At the
same time, the present $d=4$ application should not be viewed as a
general assessment of the performance of the procedure in
high-dimensional settings.

As with kernel density estimation more generally, application of the
proposed procedure in higher dimensions is subject to the curse of
dimensionality. As the dimension increases, density estimation becomes
more difficult, larger sample sizes are typically required, and the
choice of bandwidth becomes increasingly important. Computational cost
also increases because the leave-one-out density evaluations are
repeated within each bootstrap sample. The present numerical evidence
consists of the two-dimensional simulation study and a four-dimensional
real-data illustration; it should therefore not be interpreted as a
systematic assessment of moderate- or high-dimensional performance.
A systematic study of dimension-dependent finite-sample performance,
bandwidth selection, and computationally scalable implementations is an
important direction for future research.
}

\section{Discussion and Conclusion}

This paper introduced new goodness-of-fit tests based on the leave-one-out kernel density estimation for both univariate and multivariate data. The proposed procedures compare a parametric null model with a nonparametric KDE-based estimator, thereby combining the structure of parametric models with the flexibility of nonparametric methods. Several refinements were incorporated to enhance the accuracy and robustness of the approach, including boundary correction for positive-valued data and carefully chosen bandwidth selection rules in the multivariate setting. From a theoretical perspective, we established that the procedures are consistent and that, under fixed alternatives, the test statistics converge asymptotically to the Kullback–Leibler divergence, thus providing an intuitive information-theoretic interpretation of the test.

\newrev{
The simulation studies showed that the KDE-based tests maintain
reasonable type I error control across the settings considered and can
achieve high power against a range of alternatives. In several settings,
particularly under heavy-tailed, multimodal, contaminated, and
non-elliptical alternatives, the proposed procedure exhibited high
empirical rejection rates relative to some of the benchmark tests.
However, these comparisons should be interpreted in light of differences
in the null hypotheses and parameter treatment of some competing
procedures, and we do not claim that the proposed method is uniformly
superior.
}

\newrev{
Applications to real datasets further illustrated the practical use of
the proposed methodology. For the yarn breaking strength data, both the
KDE-based and classical procedures provided no evidence against the
Laplace model. For the river flow data, the KDE-based procedure with
boundary correction likewise provided no evidence against the fitted
Gamma model. In the four-dimensional \textit{Iris setosa} application,
the Gaussian-kernel KDE-based test did not reject multivariate normality,
while several competing procedures reached different conclusions. This
example illustrates that different goodness-of-fit procedures may respond
differently to the same data and motivates reporting the results as
complementary assessments rather than interpreting one procedure as
uniformly more sensitive.
}

Overall, the findings underscore the flexibility and effectiveness of KDE-based testing frameworks in addressing both classical and modern GoF problems. Unlike procedures that rely exclusively on cumulative distribution functions or moment conditions, the KDE-based approach leverages the full shape of the density, making it adaptable to a wide variety of contexts. \newrev{Future research directions include extending the methodology to censored or dependent data, systematically investigating the sensitivity of the proposed test to bandwidth selection, exploring adaptive and test-specific bandwidth selection strategies, and designing computationally efficient implementations for high-dimensional data.}

\section*{Declarations}

\noindent\textbf{Availability of data and materials.}
The datasets analyzed in this study are publicly available and are described in the manuscript. No new data were generated for this study.

\medskip
\noindent\textbf{Competing interests.}
The authors declare that they have no competing interests.

\medskip
\noindent\textbf{Authors' contributions.}
All authors contributed equally to all aspects of the work, including the conception of the study, methodological and theoretical development, analysis, writing, review, and editing. All authors read and approved the final manuscript.

\end{document}